# High-rate continuous-variable quantum key distribution over 100 km fiber with composable security

Heng Wang[1], Yang Li[1], Ting Ye[1], Li Ma[1], Yan Pan[1], Mingze Wu[2], Junhui Li[2], Yiming Bian[2], Yaodi Pi[1], Yun Shao[1], Jie Yang[1], Jinlu Liu[1], Ao Sun[1], Wei Huang[1], Stefano Pirandola[3], Yichen Zhang[2], Bingjie Xu[1]

[1] National Key Laboratory of Security Communication, Institute of Southwestern Communication, Chengdu 610041, China
[2] State Key Laboratory of Information Photonics and Optical Communications, School of Electronic Engineering, Beijing University of Posts and Telecommunications, Beijing, 100876, China
[3] Department of Computer Science, University of York, York YO10 5GH, United Kingdom

**Abstract**

Quantum key distribution (QKD), providing a way to generate secret keys with information-theoretic security, is arguably one of the most significant achievements in quantum information. The continuous-variable QKD (CV-QKD) offers the potential advantage of achieving a higher secret key rate (SKR) within a metro area, as well as being compatible with the mature telecom industry. However, the SKR and transmission distance of state-of-the-art CV-QKD systems are currently limited. Here, based on the novelly proposed orthogonal-frequency-division-multiplexing (OFDM) CV-QKD protocol, we demonstrate for the first time a high-rate multi-carrier (MC) CV-QKD with a 10 GHz symbol rate that achieves Gbps SKR within 10km and Mbps SKR over 100 km in the finite-size regime under composable security against collective attacks. The record-breaking results are achieved by suitable optimization of subcarrier number and modulation variance, well-controlled excess noise induced by both OFDM mechanism and efficient DSP scheme, and high-performance post-processing capacity realized by heterogeneous computing scheme. The composable finite-size SKR reaches 1779.45 Mbps@5km, 1025.49 Mbps@10km, 370.50 Mbps@25km, 99.93 Mbps@50km, 25.70 Mbps@75km, and 2.25 Mbps@100km, which improves the SKR by two orders of magnitude and quintuples the maximal transmission distance compared to most recently reported CV-QKD results [Nature Communications, 13, 4740 (2022)]. Interestingly, it is experimentally verified that the SKR of the proposed MC CV-QKD can approach five times larger than that of the single-carrier CV-QKD with the same symbol rate without additional hardware costs. Our work constitutes a critical step towards future high-speed quantum metropolitan and access networks.

## Introduction

Quantum key distribution (QKD) can guarantee secure keys exchange between two remote parties with information-theoretic security based on the laws of quantum physics [1-3]. QKD systems encoding information on continuous variable (CV) quadratures of coherent states and decoding by coherent receivers offer the major potential advantages of achieving higher secret key rate (SKR) within metro areas and compatibility with the mature telecom industry [4-6]. This potential has led to remarkable advancements in continuous variable QKD (CV-QKD), including protocol design [7-9], security analysis [10-16] and system implementation [17-31]. The most important key-figure-of-merit for any QKD system is to maximize the SKR over a certain distance, or conversely, to increase the distance over which a secret key can be generated. As summarized in Table 1, the SKRs and transmission distances for state-of-art CV-QKD systems significantly limit their large-scale deployments in high-speed quantum metropolitan and access networks.

To increase the SKR and transmission distance even further, a CV-QKD system needs to fulfill several key requirements. Firstly, the system must operate at a higher symbol rate (SR), e.g. 10 GHz, which in turn will result in substantial excess noises in high-speed CV quantum state preparation, transmission, and detection [33]. In particular, the fiber chromatic dispersion noise encountered over long transmission distance will significantly increase the total excess noise for high-speed CV-QKD system [34]. The SR for most state-of-art CV-QKD systems is about GHz currently [24-26], and the transmission distance for CV-QKD with 10 GHz SR is only limited to 10 km [29]. Secondly, the post-processing throughput is a strong limiting factor for real-time key generation, especially for CV-QKD system where error-correction of continuous variable raw keys with high efficiency and high throughput under ultra-low SNR is technically challenging [35-38]. The throughput of error-correction for CV-QKD with typical code rate 0.1 currently reaches 393.33 Mbps, which is far less than GHz SR [39]. To solve these problems, a multi-carrier (MC) CV-QKD scheme including 194 low-speed CV-QKD systems has been demonstrated based on wavelength-division multiplexing (WDM) method [40], but this scheme based on multiple transceivers is complex and cost ineffective. More importantly, due to the stringent demands for larger block sizes and tighter excess noise control [31], current CV-QKD systems have not yet exceeded 4.7 Mbps SKRs and extended transmission distances beyond 20 km under composable security against collective attacks, significantly limiting their practical application in high-speed secure communication.

Here, based on a novel orthogonal-frequency-division-

**Table 1** SKR comparison between the MC CV-QKD scheme and the existing QKD works. $L$: transmission distance; $R_\infty$: SKR in asymptotic regime; $R_{\text{finite}}$: finite-size SKR; SR: symbol rate; CS: composable security; PP: post-processing.

| Ref. | Attacks | SR | Modulation | $L$ (km) | Loss (dB) | $R_\infty$ (Mbps) | Block Size | $R_{\text{finite}}$ (Mbps) | CS | PP |
|---|---|---|---|---|---|---|---|---|---|---|
| This work | Finite-size collective | 10 GHz | Gaussian | 5 | 0.95 | 1819.32 | $10^{10}$ | 1779.45 | w/ | w/ |
|  |  |  |  | 10 | 1.8 | 1078.48 |  | 1025.49 |  |  |
|  |  |  |  | 25 | 4.75 | 374.19 |  | 370.50 |  |  |
|  |  |  |  | 50 | 9.5 | 112.96 |  | 99.93 |  |  |
|  |  |  |  | 75 | 12.8 | 34.63 |  | 25.70 |  |  |
|  |  |  |  | 100 | 15.8 | 12.58 |  | 2.25 |  |  |
| [24] | Asymptotic collective | 5 GBaud | Four states | 5 | 1 | 233.87 | - | - | w/o | w/ |
|  |  |  |  | 10 | 2 | 133.6 |  |  |  |  |
|  |  |  |  | 25 | 5 | 21.53 |  |  |  |  |
| [25] | Asymptotic collective | 2.5 GBaud | 16APSK | 25 | - | 49.02 | - | - | w/o | w/o |
|  |  |  |  | 50 |  | 11.86 |  |  |  |  |
|  |  |  |  | 80 |  | 2.11 |  |  |  |  |
| [26] | Asymptotic collective | 1GHz | Gaussian | 50 | 10 | 7.55 | - | - | w/o | w/o |
|  |  |  |  | 75 | 15 | 1.87 |  |  |  |  |
|  |  |  |  | 100 | 20 | 0.51 |  |  |  |  |
| [27] | Asymptotic collective | 500 MHz | Gaussian | 20 | 3.68 | 10.37 | - | - | w/o | w/o |
|  |  |  |  | 50 | 9.26 | 1.61 |  |  |  |  |
|  |  |  |  | 70 | 12.94 | 0.34 |  |  |  |  |
|  |  |  |  | 100 | 18.96 | 0.06 |  |  |  |  |
| [28] | Finite-size* collective | 600 MBaud | 256QAM | 9.5 | 1.9 | - | $5\times10^6$ | 91.8 | w/o | w/o |
|  |  |  |  | 25 | 4.3 |  |  | 24 |  |  |
| [29] | Finite-size* collective | 10 GBaud | 64QAM | 5 | 1 | 920 | $1.6\times10^7$ | 737 | w/o | w/o |
|  |  |  |  | 10 | 2 | 480 |  | 315 |  |  |
| [30] | Finite-size collective | 100 MHz | Gaussian | 100 | 15.4 | - | $10^9$ | 0.0254 | w/o | w/ |
| [31] | Finite-size collective | 100 MHz | Gaussian | 20 | 4 | - | $2\times10^8$ | 4.71 | w/ | w/ |
| [32] | Finite-size general | 2.5 GHz | Decoy BB84 | 10 | 2.2 | - | $10^8$ | 115.8 | w/ | w/ |

* The finite-size effect is only considered with worst-case estimators.

multiplexing (OFDM) protocol, we demonstrate for the first time a high-rate MC CV-QKD with 10 GHz SR that achieves Gbps SKR within 10 km and Mbps SKR over 100 km in finite-size regime with universal composability against collective attacks. Specifically, the 10 GHz OFDM-based MC CV-QKD contains 5 parallel subcarriers with 2 GHz SR using only one transceiver, where the inherent chromatic dispersion noise in long-distance fiber channel is naturally mitigated. Then, a precise MC phase noise compensation (PNC) scheme that integrates pilot-tone-assisted phase recovery and TS-aided time domain superimposition equalization is designed to control the excess noise to a reasonably low level. The SKR for the OFDM-based MC CV-QKD protocol is maximized by an optimal choice of subcarrier number, based on the proposed systematical excess noise model, and by a global optimization of the modulation variance for each subcarrier, under the restricted capacity of post-processing with finite reconciliation efficiency and frame error rate (FER). Furthermore, a high-performance post-processing module based on heterogeneous computing is innovatively designed to realize the capacity of secure key generation in real-time for each subcarrier, particularly utilizing shuffled belief propagation (BP) decoding on parallel multiple GPUs that achieves an error correction throughput of up to 1.6 Gbps.

By leveraging these technological advancements, the asymptotic SKR for the OFDM-based MC CV-QKD system achieves 1819.32 Mbps@5km, 1078.48 Mbps@10km, 374.19 Mbps@25km, 112.96 Mbps@50km, 34.63 Mbps@75km, and 12.58 Mbps@100km, which marks the first instance of CV-QKD achieving Gbps SKR within 10km transmission distance and ten Mbps SKR over 100 km transmission distance. Interestingly, due to the excess noise suppression mechanism introduced by the OFDM method, the SKR of the 10 GHz MC CV-QKD system can approach five times larger than that of the single-carrier (SC) CV-QKD system with the same SR without additional hardware costs. Furthermore, the composable finite-size SKR reaches 1779.45 Mbps@5km, 1025.49 Mbps@10km, 370.50 Mbps@25km, 99.93 Mbps@50km, 25.70 Mbps@75km, and 2.25 Mbps@100km, which achieves the first CV-QKD with Gbps SKR and 100km transmission distance in finite-size regime with universal composability against collective attacks to date. Compared to the stat-of-art CV-QKD results [31], our work improves the composable finite-size SKR by two orders of magnitude and quintuple the maximal transmission distance. Compared to the latest discrete variance QKD (DV-QKD) work [32], this CV-QKD work improves the SKR by one order of magnitude over 10 km metropolitan area.

## Results

**Composable secure key**

The security of our CV-QKD system is based on the composable security framework. Using our OFDM technology, Alice Gaussianly modulates $N$ independent coherent states on $N$ modes (or subcarriers). She then transmits the prepared quantum states to Bob through a lossy quantum channel. Bob measures the received quantum states with a heterodyne receiver and obtains the raw keys from each mode after digital demodulation. Each of these keys form an $N_t$-long key-string. Then, Alice and Bob perform post-processing, including the steps of reverse reconciliation, error correction (EC), parameter estimation (PE), and privacy amplification (PA) [41]. By randomly sacrificing $m$ points for PE in each subcarrier string, the effective key generation string will be reduced to $n=N_t-m$ points in each subcarrier. We adopt the formula for the composable key rate derived in [42], which is here assumed to be applied to each subcarrier independently. Considering the procedures of EC and PA on the $n$ key generation point, the length of the composable secret key will be reduced to $s_n$ which can be bounded using the leftover hash bound as [41]

$$s_n \leq H_{\min}^{\varepsilon_s}\left(B^n|E^n\right)_{\sigma^n} - leak_{EC} + \log_2\left(2\varepsilon_h^2 \varepsilon_{cor}\right) \quad (1)$$

where $H_{\min}^{\varepsilon_s}\left(B^n|E^n\right)_{\sigma^n}$ is the smooth-min entropy of the subnormalized state $\sigma^n$ of Bob's quantum system $B^n$ conditioned on Eve's quantum system $E^n$ with smoothing parameter $\varepsilon_s$. Here, $\varepsilon_h$ is the security parameter related to PA, $\varepsilon_{cor}$ is the security parameter for error verification, and $leak_{EC}$ is the information leaked during EC.

To apply the asymptotic equipartition property (AEP) and calculate the secret key length, the smooth-min entropy of the sub normalized state $\sigma^n$ after EC can be replaced by the normalized state $\rho^{\otimes n}$ before EC according to the following inequality [42]:

$$H_{\min}^{\varepsilon_s}\left(B^n|E^n\right)_{\sigma^n} \geq H_{\min}^{\varepsilon_s}\left(B^n|E^n\right)_{\rho^{\otimes n}} \quad (2)$$

Thereby, the smooth-min entropy can be bounded by von Neumann entropy via the AEP as [43]

$$H_{\min}^{\varepsilon_s}\left(B^n|E^n\right)_{\rho^{\otimes n}} \geq nH\left(B|E\right)_\rho - \sqrt{n}\Delta_{aep} \quad (3)$$

where

$$\Delta_{aep} = 4\log_2\left(\sqrt{2^d}+2\right)\sqrt{-\log_2\left(1-\sqrt{1-\varepsilon_s^2}\right)} \quad (4)$$

and $d$ is the quantization bits of the ADC in Bob's system.

The conditional von Neumann entropy can be expanded as

$$H\left(B|E\right)_\rho = H\left(l|E\right)_\rho = H(l) - \chi(l:E)_\rho \quad (5)$$

where $l$ is Bob's variable, $H(l)$ is the Shannon entropy of $l$, and $\chi(l:E)_\rho$ is Eve's Holevo bound with respect to $l$. According to [42], the asymptotic SKR can be expressed as

$$R_\infty = H(l) - \chi(l:E)_\rho - n^{-1}leak_{EC} \quad (6)$$

and the secret key length should satisfy the following bound

$$s_n \leq nR_\infty - \sqrt{n}\Delta_{aep} + \log_2\left(2\varepsilon_h^2\varepsilon_{cor}\right) \quad (7)$$

Since EC has success probability $p_{EC}$, and only a fraction $n/N_t$ of the total initial systems is used for key generation, the composable SKR under collective attacks can be upper bounded as [42]

$$R \leq R_{UB} = \frac{p_{ec}\left[nR_\infty - \sqrt{n}\Delta_{aep} + \log_2\left(2\varepsilon_h^2\varepsilon_{cor}\right)\right]}{N_t} \quad (8)$$

Using the direct part of the leftover hash bound [43], the lower bound of composable SKR with optimal PA can be calculated as [42]

$$R \geq R_{LB} = \frac{p_{ec}\left[nR_\infty - \sqrt{n}\Delta_{aep} + \log_2\left(2\varepsilon_h^2\varepsilon_{cor}\right)-1\right]}{N_t} \quad (9)$$

In experiments, $R_\infty$ should be computed as

$$R_\infty = \beta\left[I\right]_{(T,\hat{\xi})} - \left[\chi_\rho\right]_{(T_{wc},\xi_{wc})} \quad (10)$$

where $T$ and $\hat{\xi}$ are the estimated values of transmittance and excess noise, respectively. The worst-case values can be calculated accounting for the value $\varepsilon_{pe}$ of the PE security parameter [42]

$$T_{wc} \simeq T - w\sigma_T, \quad (11)$$

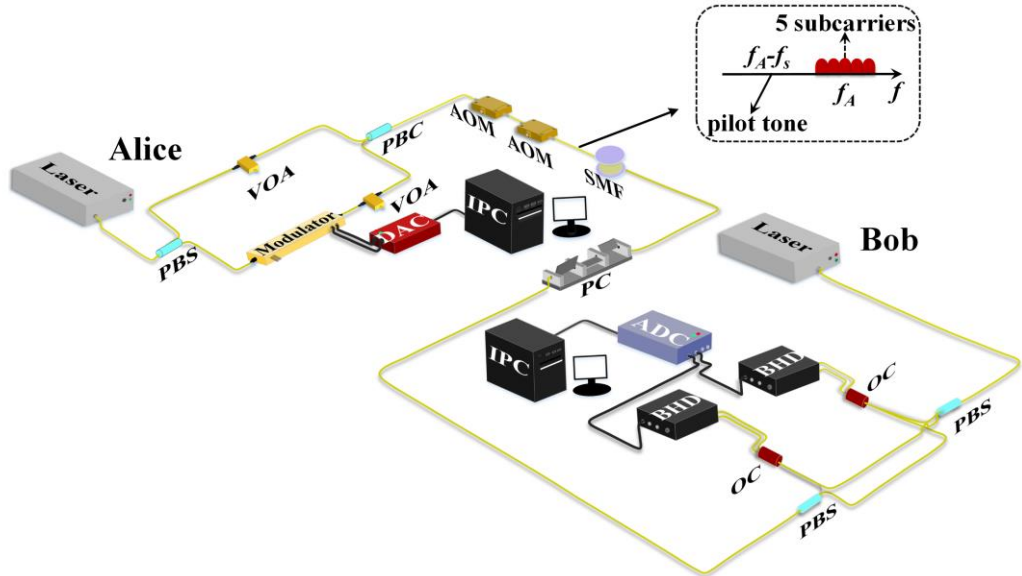

**Fig. 1** Schematic setup of the proposed OFDM-based MC CV-QKD scheme. PBS: polarization beam splitter; VOA: variable optical attenuator; DAC: digital to analog converter; PBC: polarization beam combiner; AOM: acousto-optic modulator; SMF: single mode fiber; PC: polarization controller; OC: optical coupler; BHD: balance homodyne detector; ADC: analog to digital converter; IPC: industrial personal computer.

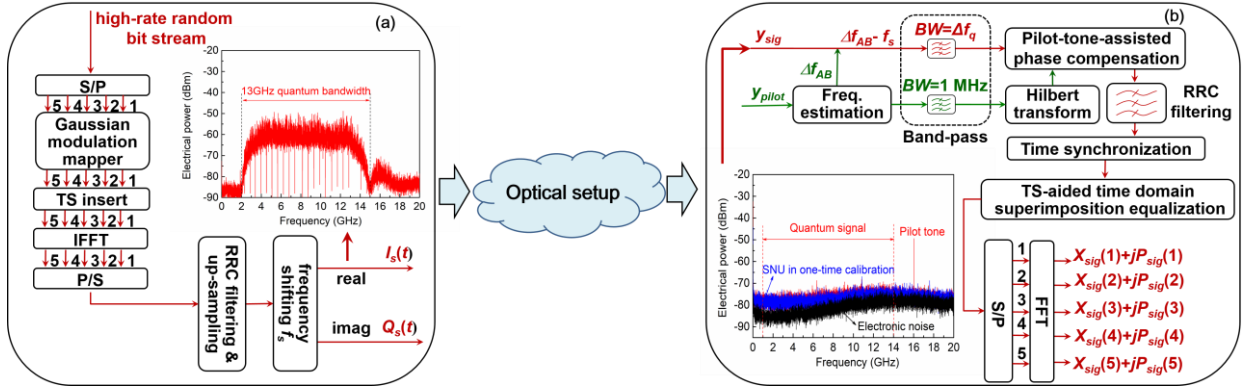

**Fig. 2** The DSP routine of the OFDM-based MC CV-QKD experiment, (a) OFDM-based MC generator, (b) OFDM-based MC demodulator. S/P: serial-to-parallel conversion; TS: training sequence; IFFT: inverse fast Fourier transform; P/S: parallel-to-serial conversion; CP: cyclic prefix; BW: bandwidth; RRC: root raised cosine; DAC: digital-to-analog converter; FFT: fast Fourier transform; SNU: shot noise unit.

$$\xi_{wc} \simeq \frac{T}{T_{wc}}\xi + w\sigma_\xi, \quad (12)$$

where

$$\sigma_T = \frac{2T}{\sqrt{2m}}\sqrt{\left(\xi + \frac{2+v_{el}}{\eta T}\right)/V}, \quad (13)$$

$$\sigma_\xi = \frac{1}{\sqrt{m}}\frac{\eta T \xi + v_{el} + 2}{\eta T_{wc}}, \quad (14)$$

and

$$w = \sqrt{2}erf^{-1}(1-2\varepsilon_{pe}), \quad (15)$$

with $erf^{-1}(\cdot)$ being the inverse error function.
Considering the steps of EC, PE and PA, the protocol has total epsilon security

$$\varepsilon = \varepsilon_{cor} + \varepsilon_s + \varepsilon_h + 2\varepsilon_{pe} \quad (16)$$

for each subcarrier. In our paper, each epsilon is set to $10^{-10}$, so we have $\varepsilon=5\times10^{-10}$ for each subcarrier under a block size.

**Experimental setup**

Using the setup shown in Fig. 1, we experimentally demonstrate the OFDM-based MC CV-QKD system with optimal $N=5$ independent subcarriers over typical transmission distances of 5 km, 10 km, 25km, 50 km, 75km, and 100km, respectively. At Alice's site, a continuous optical carrier at frequency $f_A$ from Alice's laser is split into two optical paths by a polarization beam splitter (PBS). One path is Gaussian modulated in an IQ modulator (FUJITSU FTM7962EP) by the OFDM signal $I_s$ and $Q_s$ with shifting frequency $f_s=8.5$ GHz and SR $f_{sym}=10$ GHz, which are generated from a two-channel digital to analog converter (DAC) in an arbitrary waveform generator (AWG, Keysight M8195A) with sampling rate of 30 GSa/s. Fig. 2(a) demonstrates the digital signal processing (DSP) routine of the OFDM-based MC generator at Alice's site and the desired quantum frequency spectra with 13 GHz bandwidth. The other path is directly attenuated to be a pilot tone with reasonable amplitude. The prepared MC quantum signal and pilot tone are transmitted through a SMF (ITU-T G.652) with different frequency bands and orthogonal polarization states. At Alice' site, two cascaded acousto-optic modulators (AOMs) are used as a high-speed optical switch with a 100 dB ER to realize real-time SNU calibration within short time period (see Methods for more details).

In our experiment, the modulation variance for each subcarrier can be monitored and precisely controlled with precision of 0.1 SNU. To maximize the SKR, we propose a global optimization method of modulation variance for each subcarrier under restricted capacity of post-processing with finite reconciliation efficiency and FER [44], based on which the modulation variances of the 5 subcarriers are independently optimized to be 3.8, 3.6, 3.8, 4.1, and 4.1 SNU over 50 km transmission distance as an example. The detailed modulation variance optimization routine and experimental results can be found in the Methods and Supplementary Materials.

At Bob's site, the received MC quantum signal and pilot tone are polarization-corrected by a polarization controller (PC) and then separated by a PBS. This configuration allows the MC quantum signal and pilot tone to be separately detected using two wideband balanced homodyne detectors (BHDs, Optilab BPR-23-M) with 23 GHz bandwidth. The LLO signals used for detection are independently generated from Bob's laser, which is identical to Alice's laser (NKT Photonic Basik E15). The frequency difference $\Delta f_{AB}$ between the two lasers is set to approximately 16 GHz. The output signals from the two BHDs are collected and digitized by a two-channel analog to digital converter (ADC) in a high-speed digital signal analyzer (DSA, Keysight DSAZ254A) with sampling rate 40 GSa/s. The AWG and DSA are synchronized using a 100 MHz reference clock and triggered with the switching signals of AOMs. After heterodyne detection, the measurement results of the 5 subcarriers can be extracted using the designed OFDM-based MC demodulator at Bob's site. The corresponding DSP routine and the detected frequency spectra are depicted in Fig. 2(b). It is worth noting that the SNU is measured with the one-time calibration method when Alice's laser is off and Bob's laser is on.

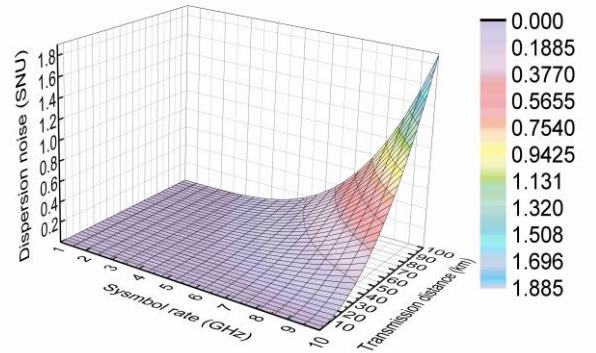

**Fig. 3** Simulated dispersion noise at different symbol rates and different transmission distances.

For the proposed MC CV-QKD, a real time post-processing with throughput larger than 1.6 GHz is required for each subcarrier with 2 GHz SR excluding the disclosed 1/5 percentage of TS. To meet this requirement, we innovatively designed a high-performance post-processing module based on heterogeneous computing, which can achieves an error correction throughput exceeding 1.6 GHz by utilizing shuffled BP decoding on parallel 6 graphics processing units (GPUs, NVIDIA Tesla A100). Moreover, a single GPU can achieve a worst-case error correction throughput of 279.2 Mbps over typical transmission distances of 5 km, 10 km, 25 km, 50 km, 75 km, and 100 km. The detailed post-processing routine and results can be found in the Methods and Supplementary Materials.

**Multi-carrier noise suppression**

It is well known that the performance of a CV-QKD system is very sensitive to its excess noise. Compared with traditional SC CV-QKD system, the MC system based on OFDM scheme will introduce new types of excess noise due to the crosstalk between different subcarriers in the process of multi-carrier quantum state modulation, transmission and detection. To evaluate the system performance and realize parameter optimization, we propose a systematical excess noise model for the OFDM-based MC CV-QKD system. The excess noise $\varepsilon_k$ of the $k$-th subcarrier can be mainly classified as [33]

$$\varepsilon_k = \varepsilon_{RIN}(k) + \varepsilon_{DAC}(k) + \varepsilon_{LE}(k) + \varepsilon_{mod}(k) + \varepsilon_{phase}(k), \quad (17)$$

where $\varepsilon_{RIN}(k)$ denotes the laser intensity noise, $\varepsilon_{DAC}(k)$ represents the quantization noise due to practical DAC, $\varepsilon_L(k)$ is the photon-leakage noise that mainly stems from the intense pilot tone, and $\varepsilon_{mod}(k)$ denotes the modulation noise of the $k$-th subcarrier that consists of finite modulation extinction ratio (ER) noise, in-phase and quadrature (IQ) imbalance noise and intermodualtion distortion (ID) noise [45]. It is remarkable that the ID noise induced in the MC quantum state modulation process cannot be ignored when $N$ is large enough. Therefore, an optimal subcarrier number $N=5$ for best SKR is chosen based on the proposed excess noise model. The detail excess noise model and the optimization of $N$ can be explained in the Methods and Supplementary Materials.

The phase noise $\varepsilon_{phase}(k)$ that dominates the excess noise mainly contains common phase error (CPE) noise and inter-carrier interference (ICI) noise. The CPE noise is mainly determined by the fiber chromatic dispersion noise, the channel phase difference and the time-varying laser phase difference. Specifically, the fiber chromatic dispersion noise encountered over long transmission distance will significantly increase for CV-QKD system with high SR, as shown in Fig.3, which can be effectively reduced by operating parallelly at lower SR with MC scheme even for long transmission distances [34]. For example, the dispersion noises for CV-QKD system with 10 GHz and 2

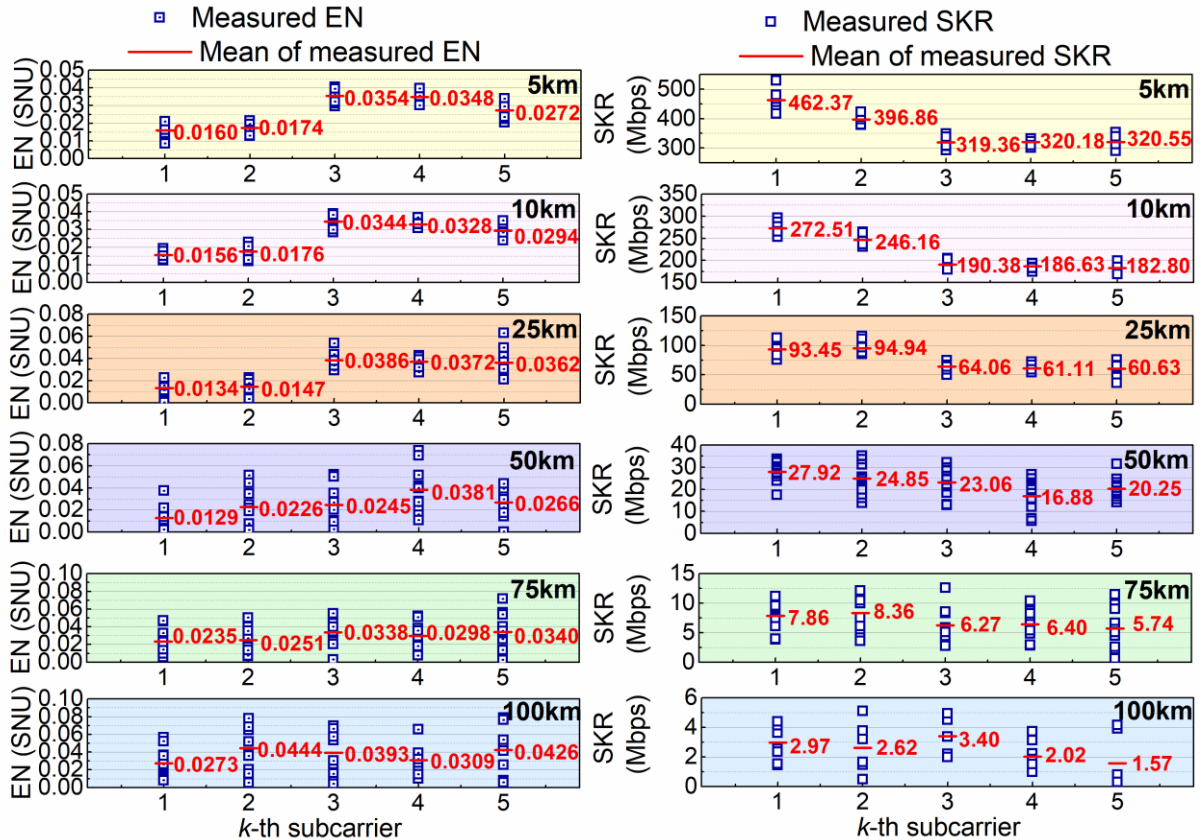

**Fig. 4** Experimentally estimated excess noises and asymptotic SKRs for 5 independent subcarriers over typical transmission distances of 5 km, 10 km, 25 km, 50 km, 75 km, and 100 km. EN: excess noise; SKR: secret key rate.

GHz SR correspond to 0.0294 and 4.711×10⁻⁵ in shot noise unit (SNU) over 50 km fiber channel, respectively. In the MC CV-QKD system, one can transform a SC quantum state with SR $f_{sym}$ into N parallel subcarrier quantum state with SR $f_{sym}/N$, which in principle can naturally mitigate the inherent chromatic dispersion noise. Furthermore, the channel phase difference can be effectively managed by our proposed PNC scheme. The ICI noise of the *k*-th subcarrier arises from the interference of other subcarriers ($i \neq k$) due to the existence of time-varying laser phase difference [46], which can be controlled to a reasonable low value for small N.

In order to realize a reasonably low excess noise, a high-precise MC PNC scheme is designed to eliminate the phase noise in the implemented setup. In this scheme, a pilot-tone-assisted PNC method is used to compensate fast-drift phase noise of the MC quantum signal which mainly originates from the rapid phase disturbances of two independent lasers and the fiber channel [47]. After the fast-drifting PNC, a TS-aided time domain superimposition equalization method is proposed for further eliminating the slow-drift phase noise with 1/5 percentage of TSs embedded in MC quantum signal. It is worth noting that the accuracy of compensating slow-drift phase noise can be significantly improved by superimposing TSs of multiple blocks for a TS with higher SNR. Importantly, this TS-aided PNC method reduces the high linearity requirements of fast data acquisition cards for high SR CV-QKD without depending on TS with high SNR. Finally, modulation noise, channel noise, and detection noise originating from the pilot tone with high signal-to-noise ratio (SNR) can be effectively minimized in our experiment. This is achieved through independent preparation, multiplexing transmission, and separate detection of the weak MC quantum signal and intense pilot tone. In particular, the spontaneous Rayleigh or Brillouin scattering spectra caused from intense pilot tone can be avoided by placing a 2GHz frequency spacing between pilot tone and MC quantum signal.

**Performance and secret key distillation**

Choosing optimal modulation variances for each subcarrier with $f_{sub}$=2 GHz, the excess noises of the 5 subcarriers are experimentally estimated with block size $1.3 \times 10^7$ over typical transmission distances of 5 km, 10 km, 25 km, 50 km, 75 km and 100 km, respectively, as shown in Fig. 4. The average asymptotic

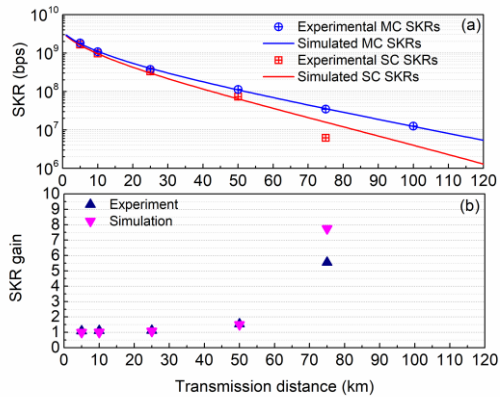

**Fig. 5** Measured SKRs comparision between MC and SC CV-QKD over different transmission distances in asymptotic regime. MC: multi-carrier; SC: single-carrier; SKR: secret key rate.

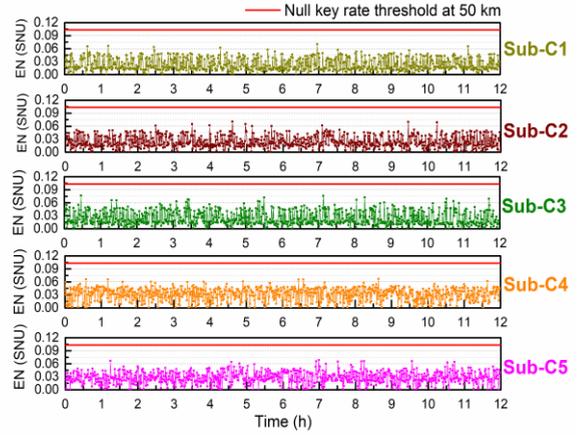

**Fig. 6** Experimental excess noises measured for 5 independent subcarriers over 12 h under the transmission distance of 50 km. EN: excess noise; Sub-C: subcarrier.

SKRs for the MC CV-QKD system are evaluated to 1819.32 Mbps@5km, 1078.48 Mbps@10km, 374.19 Mbps@25km, 112.96 Mbps@50km, 34.63 Mbps@75km, and 12.58 Mbps@100km.

To verify the potential advantage of the OFDM-based MC CV-QKD scheme, we also experimentally implement a SC CV-QKD system with 10 GHz SR under the same conditions (see Supplementary Materials for details). The numerical simulation and experimental results of SKRs for MC and SC CV-QKD setups over different transmission distance are compared in Fig. 5(a). Moreover, a SKR gain is defined as the SKR ratio between MC and SC CV-QKD protocol under the same conditions. The corresponding SKR gains at transmission distances of 5 km, 10 km, 25 km, 50km and 75km reach 1.09, 1.10, 1.13, 1.54 and 5.55, respectively, as demonstrated in Fig. 5(b). Notably, the SKR gain for 100 km cannot be provided because no secure key can be generated for the implemented SC CV-QKD setup over 100km. These results indicate the SKR of the proposed MC CV-QKD can be improved compared to that of SC CV-QKD with the same SR without additional hardware costs. Meanwhile, the SKR gains gradually increase with the transmission distance, which verifies that the MC CV-QKD scheme is more prominent for long transmission distance. Besides, the experimentally measured SKR gain is slightly lower than the theoretical estimation value over 75 km as in Fig. 5(b) because part of the dispersion phase for SC CV-QKD is compensated by our proposed PNC scheme. Nevertheless, the compensation effect of the SC dispersion phase will gradually become worse as the transmission distance increases. In contrast, MC CV-QKD systems have a lower PNC requirement due to their natural ability to mitigate inherent chromatic dispersion noise.

Subsequently, we evaluate the SKR performance of the MC CV-QKD system under a composable security framework in a finite-size regime, which is of particular interest in practical cryptographic applications. To distill positive SKR over a long transmission distance, a large block size is required for excess noise estimation and composable finite-size SKR evaluation. In our experiment, we make an average estimator of 10 block sizes for achieving the composable finite-size SKR with a total block size of $10^{10}$, where each block size is obtained by accumulating 80 frames of data in continuous time. Note that the precise accumulation of the 80 frames of data requires the excellent stability of our MC CV-QKD system. Figure 6 shows the experimentally estimated excess noise over continuous 12 hours over 50km transmission distance, verifying the continuous operation stability of our system. Under the block size of $10^{10}$, the worst-case excess noise and composable finite-size SKR are estimated over typical distances of 5km, 10km, 25km, 50km, 75km, and 100km, as summarized in Table 2. The MC excess noise and SKR in Table 2 represent the average excess noise and the sum of SKRs for the 5 subcarriers, respectively.

**Table 2** Estimated worst-case excess noises and composable finite-size SKRs with block size of $1\times10^{10}$ over different transmission distances. SKR: secret key rate; SNU: shot noise unit; Sub-C: subcarrier; MC: multi-carrier.

| Composable security | Sub-C1 | Sub-C2 | Sub-C3 | Sub-C4 | Sub-C5 | MC | Distance |
|---|---|---|---|---|---|---|---|
| Excess noise (SNU) | 0.0145 | 0.0253 | 0.0339 | 0.0311 | 0.0265 | 0.0263 | 5km |
| SKR (Mbps) | 454.04 | 370.06 | 356.78 | 280.70 | 317.87 | 1779.45 | |
| Excess noise (SNU) | 0.0158 | 0.0288 | 0.0346 | 0.0328 | 0.0274 | 0.0279 | 10km |
| SKR (Mbps) | 258.95 | 209.30 | 206.70 | 162.55 | 187.99 | 1025.49 | |
| Excess noise (SNU) | 0.0154 | 0.0157 | 0.0386 | 0.0362 | 0.0413 | 0.0294 | 25km |
| SKR (Mbps) | 87.61 | 104.13 | 66.50 | 57.37 | 54.89 | 370.50 | |
| Excess noise (SNU) | 0.0065 | 0.0064 | 0.0376 | 0.0453 | 0.0412 | 0.0274 | 50km |
| SKR (Mbps) | 27.65 | 32.31 | 16.06 | 11.06 | 12.85 | 99.93 | |
| Excess noise (SNU) | 0.0164 | 0.0195 | 0.0228 | 0.0356 | 0.0357 | 0.0260 | 75km |
| SKR (Mbps) | 6.54 | 7.47 | 5.92 | 2.81 | 2.96 | 25.70 | |
| Excess noise (SNU) | 0.0349 | 0.0103 | 0.0632 | 0.0613 | 0.0556 | 0.0451 | 100km |
| SKR (Mbps) | 0.03 | 2.21 | 8.4e-4 | 8.1e-4 | 5.3e-3 | 2.25 | |

**Discussion**

In this work, we have experimentally demonstrated a high-rate MC CV-QKD based on the novelly proposed OFDM protocol. By modeling MC excess noise, an optimal subcarrier number $N$=5 is chosen to achieve the maximal SKR. Thus, the MC CV-QKD operating 10 GHz SR is converted into 5 parallel subcarriers with 2 GHz SR using only one transceiver. In our experiment, the 5 low-speed subcarriers can be respectively realized with reasonably low excess noise based on the OFDM mechanism and well designed PNC scheme. Moreover, the modulation variance for each subcarrier is finely optimized under restricted capacity of post-processing with finite $\beta_k$ and $FER_k$. Besides, a high performance post-processing module with worst-case error correction throughput up to 1.6 Gbps is efficiently achieved by the innovatively designed shuffled BP decoding scheme on multi-GPUs, enabling the practical real-time extraction of the final secure keys for each subcarrier. Finally, the experimental asymptotic SKRs achieve 1819.32 Mbps@5km, 1078.48 Mbps@10km, 374.19 Mbps@25km, 112.96 Mbps@50km, 34.63 Mbps@75km and 12.58 Mbps@100km. Interestingly, comparing our MC CV-QKD with SC CV-QKD operating at 10 GHz SR, the SKR gain reaches up to 5 times without additional hardware costs. Furthermore, the more practical SKRs are reported as 1779.45 Mbps@5km, 1025.49 Mbps@10km, 370.50 Mbps@25km, 99.93 Mbps@50km, 25.70 Mbps@75km and 2.25 Mbps@100km under finite-size regime with universal composability against collective attacks.

Compared with the state-of-the-art QKD works shown in Table 1, our work performs the first experimental implementation of 10 GHz SR MC CV-QKD based on OFDM protocol, which offers a robust advantage in well mitigating the chromatic dispersion noise. Moreover, our work successfully develops a high-performance post-processing module capable of extracting the secure key of each subcarrier with 2 GHz SR in real time. As a result, our work firstly achieves a MC CV-QKD with finite-size SKR of Gbps within 10 km and Mbps over 100 km under composable security against collective attacks. Furthermore, our work improves the composable finite-size SKR by two orders of magnitude and increases the maximum transmission distance fourfold compared to the most state-of-the-art results [31], verifying the high-rate of the proposed MC CV-QKD. Compared to the most recent progress in DV-QKD with 110 Mbps@10km, this CV-QKD work improves the SKR by an order of magnitude over 10 km metropolitan area [32], indicating readiness for future high-rate practical CV-QKD deployment. Finally, the proposed OFDM-based MC CVQKD has a potential advantage of meeting various QKD requirements by flexibly distributing $N$ independent subcarriers with different SKRs and modulation protocols, even under a poor communication condition.

**Methods**
**OFDM-based MC DSP**
In our experiment, the DSP of the OFDM-based MC CV-QKD includes MC generation at Alice's site and MC demodulation at Bob's site. At Alice's site, the DSP routine of the OFDM-based MC generator is shown in Fig. 2(a) and described as follow. (i) A high-rate serial random bit stream is converted into 5 parallel low-rate random bit streams using serial-to-parallel conversion (S/P). Notably, the random bits are generated from a high-speed quantum random number generator. (ii) The 5 parallel bit streams are each mapped to Gaussian-modulated subcarriers and then insert 1/5 proportion of training sequences (TSs). After inverse fast Fourier transform (IFFT), the orthogonal Gaussian modulated subcarriers are converted to the desired OFDM signal through parallel-to-serial conversion (P/S). (iii) The OFDM signal is further shaped by a RRC filter with a roll-off factor of 0.3 and up-sampled to 30 GSa/s. (iv) The OFDM signal is frequency-shifted by $f_s$=8.5 GHz for the intermediate frequency (IF) detection in our scheme. (v) The real-part and imaginary-part of the formed OFDM signal are respectively converted by a two-channel DAC, and the obtained $I_s(t)$ and $Q_s(t)$ are applied onto two driving electrodes of the I/Q modulator, respectively.

At Bob's site, the DSP routine of the OFDM-based MC demodulator is depicted in Fig. 2(b) and described as follow. (i) The detected MC quantum signal and pilot tone are digitized as

$y_{sig}$ and $y_{pilot}$ by a high-speed DSA. (ii) The frequency difference $\Delta f_{AB}$ of two independent lasers is estimated to be 16 GHz by peaking the frequency spectra of the digitized pilot signal $y_{pilot}$. Thus, the central frequency $\Delta f_{AB}$-$f_s$ of the MC quantum signal is calculated to be 7.5 GHz, given the known shifting frequency $f_s$=8.5 GHz. (iii) The digitized MC quantum signal and pilot signal are bandpass filtered at the central frequencies $\Delta f_{AB}$-$f_s$ (7.5 GHz) and $\Delta f_{AB}$ (16 GHz), respectively. (iv) The bandpass filtered pilot signal is Hilbert transformed to extract the in-phase and quadrature components of the MC quantum signal. (v) In the proposed high precision PNC scheme, the fast-drift phase noise is initially compensated using the sharing phase of the pilot tone according to the pilot-tone-assisted phase compensation method. After the RRC matched filtering and time synchronization, the slow-drift phase noise is further compensated by employing the TS-aided time domain superimposition equalization method based on least mean square (LMS) algorithm. It is important to note that TSs of $M$ blocks are time-domain superimposed to create a TS with higher SNR, which is then used to compensate the phase noise of quantum signal in the $M$ blocks (e.g., $M$=16 in our experiment). (vi) The compensated MC quantum signal is serial-to-parallel converted and then fast Fourier transformed into 5-parallel quantum raw keys.

**Parameter optimization**
In the MC CV-QKD system, one can parallelly achieve multichannel key distribution and reduce the chromatic dispersion noise by increasing the subcarrier number $N$, however, which will induce incressed ID noise and ICI noise. Thus, an optimal $N$ should be chosen to achieve maximal SKR for the MC CV-QKD system. The SKR gains at different subcarrier number $N$ and different transmission distances are calculated and shown in Fig. 7, where the SKRs of MC CV-QKD and SC CV-QKD are simulated under the same condition. One can see from Fig. 7 that the optimal choice of $N$ approaches 5 for our experimental setups. Moreover, the SKR gain gradually increases with transmission distances due to the excess noise suppression effect provided by OFDM mechanism.

In practical experiment, the modulation variance $V_A(k)$ for the $k$-th subcarrier is globally optimized under restricted capacity of post-processing with finite $\beta_k$ and FER$_k$ [44]. Taking the 50km transmission distance as an example, the detailed optimization process of $V_A(k)$ for the $k$-th subcarrier is described as follows. (i) The transmittance $T$, excess noise $\varepsilon_k$, detection efficiency $\eta_k$ are experimentally calibrated and updated in real-time for $k$-th subcarrier. (ii) In a suitable modulation variance range (e.g. from 0 to 10 SNU), the corresponding SNRs and the error correction matrix $H$ are selected to determine the code rate $CR_k$. The function of $\beta_k$-$V_A(k)$ is computed as $CR_k/(0.5\log_2(1+SNR_k))$. (iii) Based on the experimental data, one can obtain a numerical relationship of FER$_k$-$\beta_k$ through curve fitting under the specific performance of data reconciliation and error correction. (iv) Using $\beta_k$-$V_A(k)$ and FER$_k$-$\beta_k$, the comprehensive function of asymptotic SKR $R_k$ on $V_A(k)$ is derived, enabling the estimation of the optimal modulation variances for the 5 subcarriers as follows: $V_A(1)$=3.8 SNU ($\beta_1$=0.9296, FER$_1$=0.0209), $V_A(2)$=3.6 SNU ($\beta_2$=0.9299, FER$_2$=0.0234), $V_A(3)$=3.8 SNU ($\beta_3$=0.9301, FER$_3$=0.0252), $V_A(4)$=4.1 SNU ($\beta_4$=0.9311, FER$_4$=0.0362), $V_A(5)$=4.1 SNU ($\beta_5$=0.9305, FER$_5$=0.0292). This optimization scheme ensures that the SKR for each subcarrier is optimal over a certain transmission distance.

**One-time and real-time SNU calibration**
In a realistic scenario, inaccuracies in SNU calibration significantly impact the estimated SKR. In our work, a one-time and real-time SNU calibration scheme is implemented to accurately evaluate the SKR of the CV-QKD setup. On one hand, the one-time SNU calibration method simplifies the calibration process by redefining the SNU as the sum of shot noise and electronic noise variances [48], which eliminates the need to separately measure electronic noise in practical experiment, thereby enhancing SNU calibration accuracy. In the one-time calibration, the electronic noise and detection efficiency of the practical BHD are both modeled as loss. Notably, the loss attributed to the electronic noise is untrusted since the electronic noise remains uncalibrated. On the other hand, a real-time SNU calibration method utilizes a high high-speed and high ER optical switch comprising two AOMs with 50 dB ER. During measurement, the sum of quantum signal and SNU variances are obtained in the front switch-on period, while the SNU variance is solely measured in the latter switch-off period. This allows for real-time calibration of SNU, considering that two SNUs are approximately equal within $ms$ period. Importantly, the SNU should be processed with the same DSP procedure as the MC quantum signal in the experiment.

**Post-processing.**
In MC CV-QKD experiment, the detailed processing routine of the designed high-performance post-processing module is described as follows. Firstly, the whole raw data of each subcarrier undergoes reverse multi-dimensional reconciliation (MR) without parameter estimation [49, 50]. After the reconciliation, the raw keys of Bob are converted into binary sequences, while the raw keys of Alice are converted into binary sequences with noises. Secondly, error correction is performed using all block sizes of each subcarrier based on multi-edge-type low density parity check (MET-LDPC) method [51, 52]. Six error correction matrices $H$ are correspondingly designed and optimized for experiments at the transmission distances of 5 km, 10 km, 25 km, 50 km, 75 km, and 100 km, with respective code rates of 0.33, 0.3, 0.18, 0.07, 0.0325 and 0.02. Finally, successful decoding at Alice's site triggers privacy amplification using a Toeplitz matrix [53, 54] to extract the final keys after performing new RMDR and optimal parameter estimation. If decoding fails, this set of data is made public for parameter estimation in subsequent QKD communication, and privacy amplification is terminated.

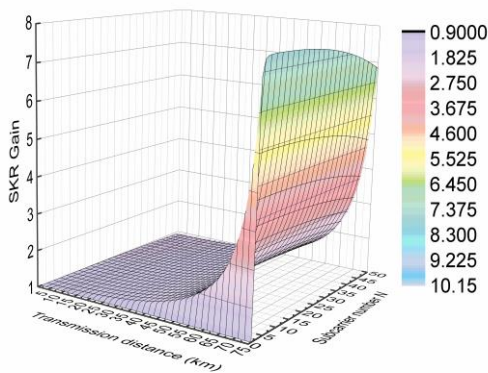

**Fig. 7** Simulated SKR gains for different subcarrier number $N$ and different transmission distances.

In the designed post-processing module, a shuffled BP decoding scheme is adopted to reduce the number of iterations and markedly increase the throughput of error correction [55]. Moreover, a cycle elimination algorithm is designed to improve the girth of the quasi-cycle MET-LDPC matrix, enabling parallel speed-up on the GPU [55]. Besides, a multiple code scheme is chosen to decrease the time of I/O operation [37]. Consequently, a single GPU can effectively achieve a worst-case error correction throughput of 279.2 Mbps over typical transmission distances of 5 km, 10 km, 25 km, 50 km, 75 km, and 100 km. The error correction utilizes 6 parallel GPUs to practically extract the final secure keys for each subcarrier in real time.

**Data availability**
All of the data that support the findings of this study are reported in the main text and Supplementary Information. Source data are available from the corresponding authors on reasonable request.

**Acknowledgments**

We acknowledge financial support from the National Key Research and Development Program of China (Grant No. 2020YFA0309704), the National Natural Science Foundation of China (Grants No. 62171418, No. 62101516, No. 62201530, and No. 62001044), the Natural Science Foundation of Sichuan Province (Grants No. 2024NSFSC0470, No. 2023NSFSC1387, No. 2023NSFSC0449, and No. 2024NSFSC0454).

**Author contributions**

W. H. and X. B. J. proposed the idea and wrote this manuscript. W. H., Y. T., P. Y. D., P. Y., and B. Y. M. carried out the experimental work. S. Y., H. W., L. J. L., Z. Y. C., and S. A. carried out the excess noise modeling. M. L., Y. J., and L. Y. carried out the post-processing work. S. P., Z. Y. C., L. J. H., and W. M. Z. carried out the theoretical analysis of the protocol. All the authors analyzed and discussed the results and contributed to writing the manuscript.

**Competing interests**

The authors declare no competing interests.